# Calculation of the zero-field splitting $|D|$ and $g_\perp$ parameters in EPR for $d^3$ spin systems in strong and moderate axial fields

Th. W. Kool[1] and B. Bollegraaf[2]

[1]Van 't Hoff Institute for Molecular Sciences, University of Amsterdam, NL 1018 WV Amsterdam

[2]University of Amsterdam

May 2010

**Abstract**

Numerical and analytical methods are used to investigate the calculation of the zero-field splitting $|2D|$ and $g_\perp$ parameters in EPR for octahedrally surrounded $d^3$ spin systems ($S = 3/2$) in strong and moderate axial crystal fields ($|D| \geq h\nu$). Exact numerical computer calculations are compared with analytical results obtained from third-order perturbation theory. From the analyses we conclude that EPR measurements performed at a single frequency with the magnetic field $H$ at a magic angle $\alpha_M$, where $62° \leq \alpha_M \leq 63°$, with respect to the axial crystal field of the $d^3$ spin system, yield an almost exact solution in third-order perturbation theory. For dual frequency experiments, i.e. X-K, X-Q and K-Q band experiments, performed with the magnetic field $H$ at an angle of $\alpha = 90°$ with respect to the axial crystal field, the ratio $\frac{h\nu}{|2D|}$ has to be smaller than 0.25 in order to use third order perturbation calculations within an error limit of 0.020% in the $g_\perp$ values. For values of $\frac{h\nu}{|2D|} \geq 0.25$ one has to proceed with exact numerical computer calculations. Finally, we conclude that for measurements performed at a single EPR frequency experiment with the magnetic field $H$ directed along two specific angles with respect to the axial crystal field of the octahedrally surrounded $d^3$ centre, i.e. $\alpha = 90°$ and $\alpha = 35°16'$ respectively, third-order perturbation theory gives non-reliable results for the $|D|$ and $g_\perp$-values.

**Introduction**

In this paper we present results of exact numerical computer and analytical third-order perturbation calculations of the zero-field splitting term $|2D|$ and $g_\perp$ for octahedrally surrounded $d^3$ systems, in the presence of strong and moderate axial fields, i.e. for $|D| \geq h\nu$.

The ground state of $d^3$ ($S = 3/2$) ions in an octahedral field is $^4A_2$ [1-3]. All excited states are lying higher in energy by amounts large compared with the spin-orbit coupling. The $^4A_2$ state is connected through spin-orbit coupling with the excited $T_{2g}$ states only [1-4]. Use of second-order perturbation theory gives a $g$-value slightly less than 2. For axially distorted (tetragonal or trigonal) octahedrally surrounded $d^3$ spin systems the following spin-Hamiltonian is used [1-3, 5]:

$$\mathcal{H} = S \cdot \bar{\bar{D}} \cdot S + \mu_B H \cdot \bar{\bar{g}} \cdot S$$

The first term represents the zero-field splitting and the second term the Zeeman interaction. The spin degeneracy of the $^4A_2$ state is partly removed into two Kramers' doublets separated by $|2D|$. The principal contribution to the $g$-shifts is caused by mixing with the excited $^4T_2$ state, which is split into an orbital singlet and an orbital doublet state. Mixing with the $^4T_2$ state, interaction with other levels and spin-spin coupling leads to the observed zero-field splitting. If the zero-field splitting $|2D|$ is much larger than the Zeeman term, only one EPR transition within the Kramers doublet with $M_S = |\pm 1/2\rangle$ levels is observed. The angular dependence of the effective $g$-values can be obtained by perturbation theory within the $^4A_2$ term using as basis the $M_s = |\pm 3/2\rangle$ and $M_s = |\pm 1/2\rangle$ wave functions and is given by:

$$\mathcal{H} = D[S_z^2 - \frac{1}{3}S(S+1)] + g_\parallel \mu_B H S_z \cos\alpha + g_\perp \mu_B H S_x \sin\alpha,$$

with α the angle between the magnetic field $H$ and the centre axis $z$.

This is justified because the excited $T_2$ states are lying much higher in energy. The effective $g$-values in first order are given by $g_\parallel^{eff} = g_\parallel$ and $g_\perp^{eff} = 2g_\perp$ with an effective spin of $S' = ½$ [1-3, 5]. The angular dependence of the effective $g$-values are altered in second- and third-order perturbation theory by [6-9]:

$$g^{eff}(\alpha) = \left(g_\parallel^2 \cos^2\alpha + 4g_\perp^2 \sin^2\alpha\right)^{\frac{1}{2}} \left[1 - \frac{3}{4}\left[\frac{g_\perp \mu_B H}{2D}\right]^2 F(\alpha)\right], \text{ where} \tag{1}$$

$$F(\alpha) = \sin^2\alpha \left[\frac{(4g_\perp^2 + 2g_\parallel^2)\sin^2\alpha - 2g_\parallel^2}{(4g_\perp^2 - g_\parallel^2)\sin^2\alpha + g_\parallel^2}\right].$$

α is the angle between the magnetic field $H$ and the axial centre axis $z$ of the $d^3$ system and $\mu_B$ is the Bohr magneton.

Hence $g^{eff}(0°) = g_\parallel$ \hfill (2)

The zero-field splitting constant $|D|$ and $g_\perp$ can be determined by measuring $g^{eff}(90°)$ at two different frequencies.

Let $\rho = \frac{g_\perp^{eff}(1)}{g_\perp^{eff}(2)}$ and $\delta = \left[\frac{H_1}{H_2}\right]^2$, then

$$g_\perp = \frac{g_\perp^{eff}(1)(\rho - \delta)}{2\rho(1-\delta)} \text{ and } |2D| = \frac{1}{2}g_\perp \mu_B \left[\frac{3(\rho H_2^2 - H_1^2)}{\rho - 1}\right]^{1/2}. \tag{3}$$

Alternatively, the spin-Hamiltonian parameters can also be determined in a single frequency experiment by using the values of $g^{eff}(90°)$ and $g^{eff}(\alpha = 35°16')$. With $g_\parallel = g_\perp$, which is almost always true in $d^3$ ($S = 3/2$) systems, the term including $F(\alpha)$ can be neglected for $\alpha = 35°16'$. We then obtain from Eqs. 1-2 [10]:

$$|2D| = g_\perp \mu_B H \left[-\frac{3}{2}\frac{g_\perp}{g^{eff}(90°)-2g_\perp}\right]^{\frac{1}{2}} \text{ and } g_\perp^2 = \frac{3}{4}g(\alpha)^2 - \frac{1}{2}g_\parallel^2. \tag{4}$$

When the zero-field splitting term is of the order of the Zeeman interaction ($|D| \approx h\nu$), one normally uses an exact computer diagonalization procedure of the secular equation. If $|D| < h\nu$, more EPR transitions can be observed and exact values for $|D|$ and $g_\perp$, within the experimental accuracy, can be obtained directly from the EPR experiment. If $|D| \geq h\nu$ only one EPR transition can be observed. $|D|$ and $g_\perp$-values can then be obtained by numerical methods or third-order perturbation theory.

In this paper we compare $|2D|$ and $g_\perp$-values obtained from third-order perturbation calculations with those obtained from exact computer diagonalization. At first, exact values for $|D|$ and $g_\perp$ will be given for different EPR frequencies, i.e. the in generally used X ($h\nu = 0.3036$ cm$^{-1}$), K ($h\nu = 0.65$ cm$^{-1}$) and Q ($h\nu = 1.17$ cm$^{-1}$) bands. These results will be compared with those obtained from third-order perturbation theory. Secondly, values for $|D|$ and $g_\perp$ obtained from exact computer diagonalization will be compared with those obtained from third-order perturbation theory in dual frequency experiments, i.e. combinations of X-K, X-Q and K-Q bands, where the magnetic field $H$ is directed at an angle of $\alpha = 90°$ with respect to the local axial crystal axis $z$ of the spin system. At last, results of exact computer diagonalization will be compared with those obtained from third-order perturbation theory in a single frequency experiment, with the magnetic field $H$ directed at two different angles, $\alpha = 90°$ and $\alpha = 35°16'$, with respect to the axial centre axis $z$.

**Exact computer diagonalization compared with third-order perturbation theory**

In calculation the $g^{eff}$-values for different values of $|D|$ with the help of an exact computer diagonalization of the secular equation, we used two computer programs. The first program was written in Fortran using routines from IMSL (International Mathematical and Statistical Library) and the second one was written in Pascal using Jacobi transformations of a symmetric matrix [11]. These values were compared with the $g^{eff}$-values obtained from third-order perturbation theory, which used the same $|D|$'s as input (see Eqs. 1-2). We took further as input data $g_\parallel = g_\perp = 2$. Deviations from $g = 2$ in axial $d^3$ systems are not so large, for instance the $g$-values for the tetragonal Fe$^{5+}$ centre in SrTiO$_3$ are $g_\parallel = 2.013$ and $g_\perp = 2.012$ [9, 12]. The following $|D|$-values, with the magnetic field $H$ directed perpendicular to the axial crystal field ($\alpha = 90°$), yielded a difference of only 0.020% or less in the $g_\perp$-values obtained from exact computer diagonalization for X, K and Q-band respectively and those obtained from third-order perturbation theory: for X-band, $|D| \geq 0.7$ cm$^{-1}$; for K-band, $|D| \geq 1.3$ cm$^{-1}$ and for Q-band, $|D| \geq 2.4$ cm$^{-1}$.

An error of only 0.020% is equal to a spread of ±0.00040 in the g-value for $g_\perp = 2$, which is very accurate in EPR experiments. In other words, if the ratio $\frac{g\mu_B H}{|2D|} = \frac{h\nu}{|2D|}$ is smaller than 0.25, the error in $g_\perp$ in third-order perturbation theory is equal to or smaller than 0.020%. Also computer diagonalization and third-order perturbation theory calculations were performed for other magnetic field directions with respect to the axial centre axis. Fig. 1 shows 0.020% error lines in the $g_\perp$-value for X, K and Q-band, respectively. The error lines are obtained from the $g^{\text{eff}}$-values of exact computer diagonalization compared with those obtained from third-order perturbation theory. For values of $|D|$ higher or equal than these lines the calculations of the exact numerical approach and those of third-order perturbation theory give the same results for $g_\perp$ within the error limit of 0.020%.

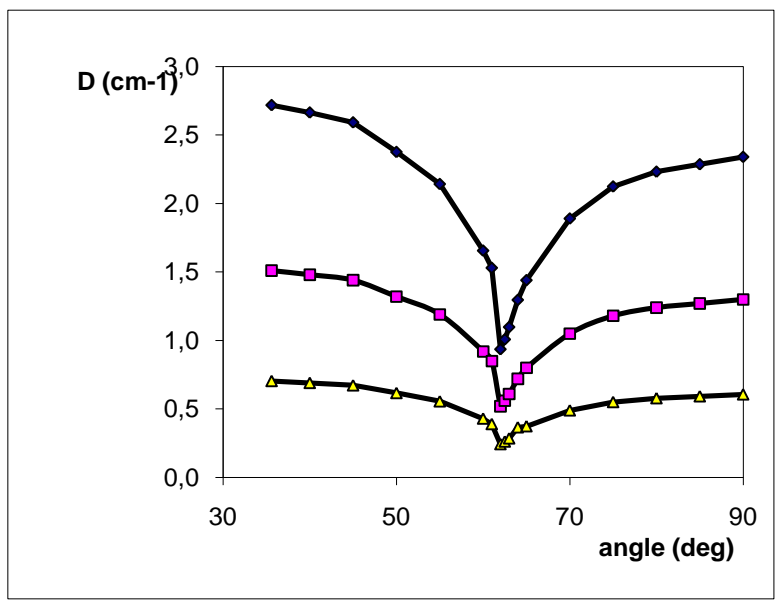

Fig. 1. The solid lines represent 0.020% error lines in the $g^{\text{eff}}$-values for X ($h\nu = 0.3036$ cm$^{-1}$) -see lower branch-, K ($h\nu = 0.65$ cm$^{-1}$) and Q ($h\nu = 1.17$ cm$^{-1}$) -see upper branch- band EPR frequencies in axial distorted octahedrally surrounded $d^3$ spin systems ($S = 3/2$). The error lines are obtained from the $g^{\text{eff}}$-values of exact computer diagonalization compared with those of third-order perturbation theory. The use of third-order perturbation calculations is allowed for $|D|$-values larger than the $|D|$-values given by these lines.

In Fig. 1 it is clearly shown that EPR measurements performed with the magnetic field $H$ oriented between $62° \leq \alpha_M \leq 63°$, with respect to the axial centre axis, give the best results in third-order perturbation theory. Fig. 1 also shows that at magic angles $62° \leq \alpha_M \leq 63°$ for $|D| < h\nu$, as said before, more transitions are observed in the EPR experiment. This means that for EPR experiments with only one transition (which can lead to more EPR lines due to different domains in a particular crystal) $|D|$ has to be larger than $h\nu$ and therefore for those systems third-order perturbation theory yields for angles $\alpha_M$ results as accurate as computer diagonalization.

Fig. 2 shows the angular dependence of the EPR spectrum with $H$ rotated in a $xz$-plane of the $d^3$ system in a single crystal for $|D| = 0.31$ cm$^{-1}$ and $h\nu = 0.3036$ cm$^{-1}$ (X-band and $|D| \approx h\nu$). The dotted line represents the first-order angular dependence of the spectrum with $g_\| = 2$ and $g_\perp^{eff} = 4$. The solid line with dots represents the angular dependence obtained from third-order perturbation theory (see Eq. 1) and the solid line represents the angular behaviour of the exact results obtained from computer diagonalization. At $\alpha_M = 62°$ the lines obtained from computer diagonalization and third-order perturbation theory cross each other. At the crossover point the results of exact computer diagonalization and third-order perturbation theory are equal. The crosses in Fig. 2 represent the error in percentage between the $g$-values obtained from computer diagonalization and third order perturbation theory. It is seen that at $\alpha_M = 62°$ the error in the $g$-value is almost 0%.

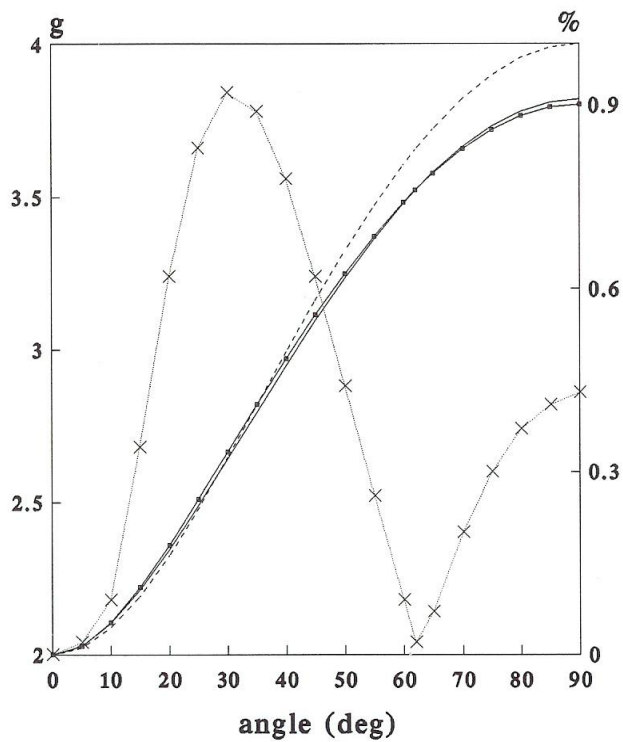

Fig. 2. Angular dependence of axial distorted octahedrally surrounded $d^3$ systems, with the magnetic field $H$ rotated in a $xz$-plane ($|D| = 0.31$ cm$^{-1}$ and $h\upsilon = 0.3036$ cm$^{-1}$). The dotted line represents first-order angular dependence. The solid line with dots represents the angular dependence of the spectrum calculated with the help of third order-perturbation theory. The solid line represents the angular behaviour of results obtained from exact computer diagonalization. The crosses give the error in percentage ($dg\%$) of the $g^{eff}$-values obtained from exact computer diagonalization and compared with the $g^{eff}$-values obtained from third-order perturbation theory. At 62° third-order perturbation theory yields the same results as an exact computer diagonalization procedure. At 35°16′, the $g^{eff}$-value of the first-order spectrum is equal to the $g^{eff}$-value obtained from third-order perturbation theory.

**Dual frequency experiment for D ≥ hυ**

As mentioned in the introduction $|D|$ and $g_\perp$ are often determined by measuring $g^{eff}(90^o)$ at two different frequencies [6-9, 12]. We are now interested in the relation between $|D|$ and $h\nu$ and want to know for which values of $|D|$ third-order perturbation theory gives reliable results with respect to exact computer diagonalization. We analysed the three well-known combinations often used in EPR experiments, i.e. X-K, X-Q and K-Q bands. The calculation procedure was as follows: first we calculated for different $|D|$-values ($g_\parallel = g_\perp = 2$), with the help of exact computer diagonalization, the exact magnetic field values. Then we put the calculated magnetic field values in Eq. 3. At last we compared the calculated $|D|$ and $g_\perp$ values obtained from Eq. 3 with the input data.

Again we used an error of only 0.020% in the $g_\perp$-values between the results of an exact computer diagonalization procedure and third-order perturbation theory. This yielded an error of 2.15% in the $|D|$-value in an X-K band experiment, an error of 4.66% in an X-Q band experiment and an error of 2.39% in a K-Q band experiment. In conclusion, the best results in a dual frequency experiment with the help of third-order perturbation theory are obtained by a dual X-K frequency EPR experiment for $|D|$-values not less than 1 cm$^{-1}$.

**Single frequency experiment for D ≥ hυ**

It was mentioned before that it is also possible to use a single frequency experiment in EPR with $H$ taken at $\alpha = 90^o$ and $\alpha = 35^o16'$, with respect to the axial centre axis $z$ [10]. For $g_\parallel = g_\perp$ the term $F(35^o16')$ in Eq. 1 becomes zero and therefore $|D|$ and $g_\perp$ can be calculated in third-order perturbation theory with the help of Eq. 4.
As shown in Fig. 2 the angular dependence of the EPR spectrum in first-order approximation (dotted line) crosses the spectrum with angular dependence calculated with third-order perturbation theory (solid line with dots) at $\alpha = 35^o16'$. This means that the $F(\alpha)$ term in Eq. 1 becomes zero at this angle.
Again the calculation procedure was as follows: an exact computer diagonalization ($g_\parallel = g_\perp = 2$) yielded $|D|$-values for α equal to $90^o$ and $35^o16'$. With the help of Eq. 4 we calculated $|D|$ and $g_\perp$ for third-order perturbation theory. The calculations, performed with an error of 0.020% in the $g_\perp$-value, yield an error of about 100% in the $|D|$-value for X, K as well as Q-band experiments. Therefore, we conclude that in a single frequency experiment with $|D| \geq h\nu$ taken at two different angles ($\alpha = 90^o$ and $\alpha = 35^o16'$) third-order perturbation theory gives non-reliable results for calculating $|D|$ and $g_\perp$ values.

**Discussion of some axial d³ (S = 3/2) systems**

We will now consider four $d^3$ centres in strong and moderate axial fields. At first we discuss the Mo$^{3+}$ impurity ion in Al$_2$O$_3$, which substitutes for the Al$^{3+}$ ion surrounded by six O$^{2-}$ ions in a trigonal crystal field [13]. The spin-Hamiltonian is given by Eq. 1 and is characteristic of a distorted octahedron with a trigonal field. For molybdenum compounds with $d^3$ configuration the zero-field splitting is larger than the Zeeman term due to the combined effect of spin-orbit coupling and trigonal field. It is now possible to use a new effective spin-Hamiltonian for the $M_s = |\pm 1/2\rangle$ doublet as described in the introduction. The EPR experiments were performed in a single frequency experiment ($h\nu = 0.3085$ cm$^{-1}$), with the magnetic field $H$ directed along two different angles ($\alpha = 90°$ and $\alpha = 35°16'$). The following EPR parameters were obtained: $D = -0.8 \pm 0.3$ cm$^{-1}$ and $g_\perp = 1.98 \pm 0.01$. The error in $|D|$ is equal to 37.5% and the one in $g_\perp$ is equal to 0.50%. According to our calculations an error of 0.10% in $g_\perp$ yields already an error of more than 100% in the $|D|$-value. Therefore, our conclusion is that the estimated value of $|D|$ is wrong. Because a single transition has been observed in the EPR experiment we can conclude that $|D| > 0.3085$ cm$^{-1}$. Therefore, it is better to analyse the Mo$^{3+}$:Al$_2$O$_3$ system again to obtain new and better $D$ and $g_\perp$-values.

The second system is the Fe$^{5+}$ impurity ion in BaTiO$_3$ substituting for the Ti$^{4+}$ ion in a trigonal field stemming from a nearby Ba$^{2+}$ vacancy [14, 15]. Later investigations with the help of externally applied uniaxial stress revealed that the centre is off-centred in one of the <111> directions [16, 17] Again the zero-field splitting is larger than the Zeeman term. Therefore, the effective spin-Hamiltonian formalism for the $M_s = |\pm 1/2\rangle$ doublet can be used. The values for $|D| = 1.15 \pm 0.05$ cm$^{-1}$ and $g_\perp = 2.013 \pm 0.003$ were obtained from dual frequency experiments (X-Q band). The error in $|D|$ is equal to 4.35% and the one in $g_\perp$ is equal to 0.15%. According to our calculations an error of 5.75% in $|D|$ yields a minimum error of 0.040% in the $g_\perp$-value. Therefore, in this system third-order perturbation theory in a dual EPR X-Q frequency experiment is allowed.

As third system we discuss the tetragonal Fe$^{5+}$ centre in SrTiO$_3$ [9, 12]. The Fe$^{5+}$ ion substitutes for a Ti$^{4+}$ ion with probably a Ti$^{4+}$ vacancy in an adjacent oxygen octahedron. Here $|D| = 0.275$ cm$^{-1}$, which is smaller than $h\nu$ (0.3036 cm$^{-1}$). This means that more EPR transitions are observable, which indeed have been observed.

At last we want to mention that up till now only in one axial $d^3$ ($S = 3/2$) system (Cr$^{3+}$-V$_o$:WO$_3$) an oxygen vacancy in the first octahedron has been found [17-19]. This leads to a high $|D|$-value of about 2.4-2.8 cm$^{-1}$. The calculation of the $|D|$-value in this system was made with the help of a computer diagonalization procedure. With the obtained value of $|D|$ the angular dependent EPR spectrum has been fitted. In general axial $d^3$ systems give lower $|D|$-values than axial $d^5$ ($S = 5/2$) systems, because in the latter more vacancies in the surrounding octahedron of the substituting impurity ion have been found.

**Conclusion**

In conclusion we found that it is not necessary to make use of exact computer diagonalization in calculating $|D|$ and $g_\perp$-values for octahedrally surrounded $d^3$ systems ($S = 3/2$) in strong and moderate axial crystal fields. By measuring the magnetic field $H$ at different angles $\alpha_M$ in a single frequency EPR experiment, where $62^o \leq \alpha_M \leq 63^o$, with respect to the axial centre axis, third-order perturbation theory is applicable within the experimental accuracy of the EPR experiment and yields the same results for the calculation of $|D|$ and $g_\perp$ as those obtained by an exact computer diagonalization procedure. Also analyses were made for dual frequency EPR experiments measured with the magnetic field **H** directed at an angle of $90^o$ with respect to the axial crystal field. We found that in dual frequency experiments the ratio $\frac{h\nu}{|2D|}$ has to be smaller than 0.25 in order to use third-order perturbation theory calculations, within an error limit of 0.020% in the $g_\perp$-value (in comparison with exact numerical calculations). In other words the zero-field splitting parameter $|D|$ has to be equal or larger than twice the Zeeman term. We also concluded that EPR measurements at a single frequency, with the magnetic field $H$ at two different angles, α = $90^o$ and α = $35^o16'$, is of no value for obtaining reliable $|D|$ and $g_\perp$ values.